\def\twocolumns{}

\ifdefined\twocolumns
  \documentclass[a4paper]{IEEEtran}
\else
  \documentclass{article}
  \usepackage{arxiv}
\fi

\usepackage[affil-it]{authblk}
\usepackage[utf8]{inputenc} 
\usepackage[english]{babel}
\usepackage[T1]{fontenc}    

\usepackage[colorlinks=true]{hyperref}       
\usepackage{url}                
\usepackage{graphicx}
\usepackage{subcaption}
\usepackage{doi}

\usepackage{amsthm,amsmath,amscd}
\usepackage{amsfonts,amssymb}
\usepackage{mathtools, cuted}

\usepackage[round]{natbib}

\usepackage[dvipsnames]{xcolor}

\definecolor{linkcolor}{rgb}{0.08, 0.38, 0.74}
\definecolor{citecolor}{rgb}{0.18, 0.55, 0.34}
\definecolor{urlcolor}{rgb}{0.03, 0.57, 0.82}

\hypersetup{
    linktocpage=true,           
    colorlinks,                 
    linkcolor={linkcolor},      
    citecolor={citecolor},      
    urlcolor={urlcolor},        
}

\newcommand*{\email}[1]{
  \href{mailto:#1}{\texttt{#1}}
}

\def\keywordname{{\bfseries \emph{Keywords}}}%
\def\keywords#1{\par\addvspace\medskipamount{\rightskip=0pt plus1cm
\def\and{\ifhmode\unskip\nobreak\fi\ $\cdot$
}\noindent\keywordname\enspace\ignorespaces#1\par}}

\newcommand{\MS}{\mbox{$M_{\odot}$}}

\newcommand{\RS}{\mbox{$R_{\odot}$}}

\title{Opacity of Ejecta in Calculations of Supernova Light Curves}

\author[1,2,4*]{M.Sh.~Potashov}
\author[1,2,3,4,**]{S.I.~Blinnikov}
\author[2,5]{E.I.~Sorokina}
\affil[1]{Keldysh Institute of Applied Mathematics, Russian Academy of Sciences, 125047, Moscow, Russia}
\affil[2]{NRC ``Kurchatov Institute'', 117218, Moscow, Russia}
\affil[3]{Dukhov Research Institute of Automatics, 127055, Moscow, Russia}
\affil[4]{Novosibirsk State University, 630090, Novosibirsk, Russia}
\affil[5]{Sternberg Astronomical Institute, Moscow State University, 119234, Moscow, Russia}
\affil[*]{\email{Marat.Potashov@gmail.com}}
\affil[**]{\email{Sergei.Blinnikov@gmail.com}}

\date{} 					

\ifdefined\twocolumns
  \usepackage{fancyhdr}
\fi

\hypersetup{
pdftitle={Opacity of Ejecta in Calculations of Supernova Light Curves},
pdfsubject={Expansion opacity},
pdfauthor={M.Sh.~Potashov, S.I.~Blinnikov, E.I.~Sorokina},
pdfkeywords={supernovae, expansion opacity, light curves, radiative transfer},
}

\begin{document}
\maketitle

\begin{abstract}
  The plasma opacity in stars depends mainly on the
    local state of matter (the density, temperature,
    and chemical composition at the point of interest), but in supernova ejecta it also depends
    on the expansion velocity gradient, because the Doppler effect shifts
    the spectral lines differently in different ejecta layers.
    This effect is known in the literature as the expansion opacity.
    The existing approaches to the inclusion of this effect, in some cases, predict
    different results in identical conditions. In this paper we compare the
    approaches of \citet{Blinnikov1996} and
    \citet{FriendCastor1983}--\citet{EastmanPinto1993} to calculating the opacity in supernova ejecta and give
    examples of the influence of different approximations on the model light
    curves of supernovae.
\end{abstract}



\section*{Introduction}
  \noindent
  \label{sec:introduction}

Radiative transfer is of paramount importance in
  a high-temperature plasma.
It is especially important in astrophysics, because most
  of the data on the Universe are obtained from radiation.
The optical plasma properties determine the interaction of matter
  with radiation and are an important part of any problem
  with radiative transfer.
The interaction of matter with radiation is characterized
  by the opacity (the reciprocal of the mean free path).
The opacity itself gives an idea of the atomic and ionic structure of materials.
In some cases, bound--bound transitions in ions create a thick
  ``forest'' of spectral lines that contribute significantly to the opacity.
It is not easy to calculate this contribution, and additional physical effects
  (for example, flow inhomogeneity, non--LTE)
  can make this calculation more difficult.
In moving matter all line frequencies are shifted due to the Doppler effect.
The light emitted in the rest frame interacts with a moving
  plasma with the absorptivity calculated
  at the shifted frequency \( \Delta \nu \).
If the matter moves nonuniformly (the velocity changes from
  point to point), then the shifts  \( \Delta \nu \) become
  dependent on the position and angle.
Thus, it is necessary to sum the contributions
  of different lines at different points on the path of the light in the plasma.

By the ``velocity gradient'' we mean the spatial derivative
  of the velocity component along the general expansion direction
  -- for a radial flow this is
\begin{equation} \label{eq:gradu}
  \frac{d u}{d r} \equiv \frac{\partial v_{r}}{\partial r}.
\end{equation}
In this paper we will assume that all formulas were
  derived at the stage of free expansion with kinematics
  \(v = r/t\), i.e., the velocity gradient is equal
  to the reciprocal of the time after explosion.
The plasma opacity can depend on the velocity gradient,
  because the Doppler effect shifts the spectral lines
  differently in different layers.
For the cases where there are many bound--bound transitions,
  i.e., a large number of lines contribute to the opacity,
  the latter is enhanced when the plasma expands with
  a nonuniform velocity field.
The expansion opacity approximation, whose interpretation
  still remains debatable, was introduced to perform
  calculations in such situations.
The problem of a proper approximation to describing
  the absorption and scattering of
  radiation in a plasma moving with a velocity gradient
  was considered in a number of papers.
Several approaches to calculating the expansion opacity
  are described in the well-known book by
  \citet{Castor2004book},
  in particular, those from
  \citet{FriendCastor1983},
  \citet{EastmanPinto1993},
  \citet{Blinnikov1996},
  \citet{BaronHauschildtMezzacappa1996},
  and
  \citet{WehrseBaschekvonWaldenfels2003}.
A puzzling fact is that in some situations the photon
  mean free paths calculated by different methods differ
  by orders of magnitude.
In this paper we will check the differences
  for the expansion opacity from
  \citet{FriendCastor1983}
  and
  \citet{EastmanPinto1993},
  on the one hand, and
  \citet{Blinnikov1996},
  on the other hand.
Below we will denote the models computed using
  the approach of
  \citet{FriendCastor1983}
  and
  \citet{EastmanPinto1993}
  by the index \texttt{E},
  because it is based in part on heuristic considerations.
The second, Blinnikov's approach will be denoted
  by the index \texttt{H},
  corresponding to the Hilbert expansion used in
  \citet{Blinnikov1996}.

We will also consider the influence of different opacity parameters
  on the predicted light curves of supernovae (SNe),
  other things being equal.


\section*{The Limiting Cases Of Strong And Weak Lines}
  \label{sec:limits}

In this section we will show that the expressions
  for the expansion opacity derived by
  \citet{Blinnikov1996}
  are reduced to the Friend--Castor (and Eastman--Pinto)
  formulas in two limiting cases:
  when all lines are strong and when all lines are weak.
We will adopt the following notation:
  \(\chi\) is the opacity (the reciprocal of the mean free path),
  \(s\equiv ct\chi\) is the expansion opacity parameter,
  where \(t\) is the time from SN explosion
  (already at the homologous stage).
Let us write the expression from
  \citet{Blinnikov1996}
  as
\makeatletter
\if@twocolumn
\begin{equation} \label{eq:sumfullpr}
  \begin{multlined}
    \chi_{\exp}^{-1}(\nu) =
      \chi_{N_\nu-1}^{-1}
      \bigl[1 - e^{-s_{N_\nu-1}(1-\nu/\nu_{N_{\nu}})} \bigr] + \\
        + \sum_{i=N_\nu}^{N_{\max}}\chi_i^{-1}
        \bigl[1 - e^{-s_i(\nu/\nu_i-\nu/\nu_{i+1})} \bigr] \\
        \exp\Bigl\{
          -\sum_{j=N_\nu}^i\Bigl[ s_{j-1}\Bigl({\frac{\nu}{\nu_{j-1}}}
          - {\frac{\nu}{\nu_j}}\Bigr)
          + {\tau_j \frac{\nu}{\nu_j}} \Bigr]
          \Bigr\} \; .
  \end{multlined}
\end{equation}
\else
\begin{equation} \label{eq:sumfullpr}
  \begin{multlined}
    \chi_{\exp}^{-1}(\nu) =
      \chi_{N_\nu-1}^{-1}
      \bigl[1 - e^{-s_{N_\nu-1}(1-\nu/\nu_{N_{\nu}})} \bigr] + \\
        + \sum_{i=N_\nu}^{N_{\max}}\chi_i^{-1}
        \bigl[1 - e^{-s_i(\nu/\nu_i-\nu/\nu_{i+1})} \bigr]
        \exp\Bigl\{
          -\sum_{j=N_\nu}^i\Bigl[ s_{j-1}\Bigl({\frac{\nu}{\nu_{j-1}}}
          - {\frac{\nu}{\nu_j}}\Bigr)
          + {\tau_j \frac{\nu}{\nu_j}} \Bigr]
          \Bigr\} \; .
  \end{multlined}
\end{equation}
\fi
\makeatother
Here, \(N_\nu\) is the number of the first line from a specified list
  that can affect the observer due to the expansion redshift,
  and \(\chi_i\) is the mean opacity in the continuum
  (or quasi-continuum) between adjacent lines \(\nu_i\) and \(\nu_{i+1}\).
The continuum can be formed through
  free--bound and
  free--free transitions,
  electron scattering,
  and a superimposed quasi-continuum formed by
  a ``forest'' of lines
  through various expansion mechanisms in the rest frame.
The values of the parameter \(s_i\)
  between the lines can differ due to the differences
  of \(\chi_i\) in the continuum.
In the last sum over \(j\) \(\nu_{j-1}\) is assumed
  to be equal to \(\nu\) at \(j=N_\nu\).
It is convenient to rewrite Eq.~(\ref{eq:sumfullpr})
  partly via the wavelength \(\lambda=c/\nu\)
  by introducing \(\delta\lambda_i \equiv \lambda_{i}-\lambda_{i+1}\):
\makeatletter
\if@twocolumn
\begin{equation} \label{eq:sumdlam}
  \begin{multlined}
    \chi_{\exp}^{-1}(\nu) \approx
    \sum_{i=N_\nu-1}^{N_{\max}}\chi_i^{-1}
    \bigl[1 - e^{-s_i(\delta\lambda_i/\lambda)} \bigr] \\
    \exp\Bigl\{
      -\sum_{j=N_\nu}^i\Bigl[ s_{i-1} \Bigl({\frac{\delta\lambda_j}{\lambda}}\Bigr)
      + {\frac{\tau_j\lambda_i}{\lambda}} \Bigr]\Bigr\} \; .
  \end{multlined}
\end{equation}
\else
\begin{equation} \label{eq:sumdlam}
  \begin{multlined}
    \chi_{\exp}^{-1}(\nu) \approx
    \sum_{i=N_\nu-1}^{N_{\max}}\chi_i^{-1}
    \bigl[1 - e^{-s_i(\delta\lambda_i/\lambda)} \bigr]
    \exp\Bigl\{
      -\sum_{j=N_\nu}^i\Bigl[ s_{i-1} \Bigl({\frac{\delta\lambda_j}{\lambda}}\Bigr)
      + {\frac{\tau_j\lambda_i}{\lambda}} \Bigr]\Bigr\} \; .
  \end{multlined}
\end{equation}
\fi
\makeatother
Let there be many lines, i.e., \(\delta\nu_i \ll \nu\),
  where \(\delta\nu_i \equiv \nu_{i+1}-\nu_i\)
  (in practice, \(\delta\nu_i/\nu\) can be \(\sim 10^{-6}\),
  for example, for iron lines),
  and we will assume that the entire range important
  for the effect is \(\Delta\nu << \nu\).
This inequality is not so strong as the previous one,
  because either \(\Delta\nu = \nu/s\) or \(\Delta\nu \sim \nu(u_{\max}/c)\).
The stronger of these inequalities must hold:
  at small \(s\) it is clear that the Doppler effect
  ceases to work at \(\Delta\nu > \nu(u_{\max}/c)\).
We will then get
\makeatletter
\if@twocolumn
\begin{equation} \label{eq:sumdnupr}
  \begin{multlined}
    \chi_{\exp}^{-1}(\nu) \approx
    \sum_{i=N_\nu-1}^{N_{\max}}\chi_i^{-1}
    \bigl[1 - e^{-s_i(\delta\nu_i/\nu)} \bigr] \\
    \exp\Bigl\{
      -\sum_{j=N_\nu}^i\Bigl[ s_{i-1} \Bigl({\frac{\delta\nu_j}{\nu}}\Bigr)
      + \tau_j \Bigr]\Bigr\} \; .
  \end{multlined}
\end{equation}
\else
\begin{equation} \label{eq:sumdnupr}
  \begin{multlined}
    \chi_{\exp}^{-1}(\nu) \approx
    \sum_{i=N_\nu-1}^{N_{\max}}\chi_i^{-1}
    \bigl[1 - e^{-s_i(\delta\nu_i/\nu)} \bigr]
    \exp\Bigl\{
      -\sum_{j=N_\nu}^i\Bigl[ s_{i-1} \Bigl({\frac{\delta\nu_j}{\nu}}\Bigr)
      + \tau_j \Bigr]\Bigr\} \; .
  \end{multlined}
\end{equation}
\fi
\makeatother
Note that, in contrast to the last expression,
  in Eq.~(\ref{eq:sumdlam}) there was no need to assume
  that \(\delta\lambda_i << \lambda\),
  although this condition is almost always fulfilled in practice.
If \(s_i(\delta\nu_i/\nu)\) is great, i.e.,
  the lines are few, and the parameter \(s_i\) is great,
  then all of the exponentials in~(\ref{eq:sumdnupr})
  are small and the expansion effect vanishes:
  \(\chi_{\exp}=\chi_{N_{\nu} - 1}\).
The case where the parameter \(s_i\) is not
  too great, say,
  \(s_i < 10^3\) and \(\delta\nu_i/\nu \sim 10^{-6}\) is less trivial.
Then, \(s_i(\delta\nu_i/\nu)\) is small
  and the first exponential can be expanded:
  \(1 - \exp[-s_i(\delta\nu_i/\nu)]=s_i(\delta\nu_i/\nu)\).
However, since \(s_i=\chi_i ct\), we have
\begin{equation} \label{eq:ctnupr}
  \chi_{\exp}^{-1}(\nu) \approx ct
  \sum_{i=N_\nu-1}^{N_{\max}}
  {\frac{\delta\nu_i}{\nu}}
  \exp\Bigl\{
    -\sum_{j=N_\nu}^i\Bigl[ s_{i-1} \Bigl({\frac{\delta\nu_j}{\nu}}\Bigr)
    + \tau_j \Bigr]\Bigr\} \; .
\end{equation}
The same is trivially rewritten via the wavelengths:
\begin{equation} \label{eq:ctlampr}
  \chi_{\exp}^{-1}(\nu) \approx ct
  \sum_{i=N_\nu-1}^{N_{\max}}
  {\frac{\delta\lambda_i}{\lambda}}
  \exp\Bigl\{
    -\sum_{j=N_\nu}^i\Bigl[ s_{i-1} \Bigl({\frac{\delta\lambda_j}{\lambda}}\Bigr)
    + \tau_j \Bigr]\Bigr\} \; .
\end{equation}
Let us compare our formulas with the approximation
  for the expansion opacity proposed
  by
  \citet{FriendCastor1983}
  and
  \citet{EastmanPinto1993}
  (below we will designate the reference to this paper as \texttt{E}).
In this approximation the contribution of the lines to
  the opacity in a given frequency interval \((\nu,\nu+\Delta\nu)\)
  is assumed to be given in the case of homologous expansion
  by the expression
\begin{equation} \label{eq:epopac}
  \chi_{\rm E}={\frac{\nu}{\Delta\nu}}{\frac{1}{ct}}\sum_j
  \left\{1-\exp\left[-\tau_j\right]\right\}\; ,
\end{equation}
where the sum is taken over all lines in the interval \((\nu,\nu+\Delta\nu)\)
  and \(\tau_j\) is the Sobolev optical depth
  in line \(j\)
  \citep{Sobolev1947book, Sobolev1960book}:
\begin{equation}
  \tau_{j}(r)={\frac{hc}{4\pi}}{
    \frac{(n_l B_{lu}-n_u B_{ul})}{(\partial v/\partial r)}} \; .
  \label{sobtau}
\end{equation}

In principle, Eq.~(\ref{eq:epopac}) has a slightly different meaning
  than our \(\chi_{{\exp}\nu}\) -- we obtain the ``monochromatic''
  \(\chi_\nu\) at frequency \(\nu\),
  while Friend--Castor and \texttt{E} obtain the mean
  in an interval \(\Delta\nu\).
Therefore, for comparison with \texttt{E},
  we additionally need to average our \(\chi_{{\exp}\nu}\)
  over the interval \(\Delta\nu\).
Let us define the mean over the interval \(\Delta\nu\)
  as the mean free path:
\begin{equation} \label{eq:chimean}
  \frac{1}{\chi_\mathrm{H}} =
    \frac{1}{{\Delta\nu}}
    \int \limits_{\Delta\nu} \frac{\mathrm{~d} \nu}{\chi_{\nu \exp}}.
\end{equation}

The \texttt{E} result was obtained from the following
  simple heuristic considerations.
The \texttt{E} opacity or rather the mean extinction coefficient
  in an interval \((\nu,\nu+\Delta\nu)\) is the mean number
  of interactions between a photon and lines as they
  are Doppler shifted by \(\Delta\nu\) divided by
  the distance traveled \(\sim ct\Delta\nu/\nu\).
One would think, on the first impression,
  that Eq.~(\ref{eq:epopac}) cannot be derived from
  our Eq.~(\ref{eq:ctnupr}), because the exponentials of \(-\tau_j\)
  enter into both expressions,
  but for the mean free path in our case and for
  the reciprocal of the mean free path in \texttt{E}.
In fact, the estimates of the mean opacities from~(\ref{eq:ctnupr})
  agree with~(\ref{eq:epopac}).

From~(\ref{eq:ctnupr}) in the case of strong lines,
  i.e., lines with a Sobolev optical depth \(\tau_j>1\),
  we find that the sum does not go up to \(N_{\max}\),
  but is truncated at the first \(k\),
  such that \(\tau_k>1\) and
  \(\sum_{i=N_\nu}^{k-1}\tau_i<1\).
Then, we have the following estimate:
\begin{equation} \label{eq:ctnuest}
  \chi_{\exp}^{-1}(\nu)\approx
  ct{\frac{\nu_k - \nu}{\nu}} \approx
  ct {\frac{\Delta\nu}{N_{\rm strong}\nu}} \; ,
\end{equation}
  where \(N_{\rm strong}\) is the number of strong lines in the
    interval \(\Delta\nu\).
We see that this coincides with \texttt{E},
  because mostly \(N_{\rm strong}\) strong lines
  with a Sobolev optical depth greater than
  unity contribute to the sum in~(\ref{eq:epopac}).

If there are no strong lines in the interval,
  then the result of averaging~(\ref{eq:ctnupr})
  also coincides with \texttt{E} in the case of weak lines.
Let all lines in the interval \(\Delta\nu\) have
  a small optical depth, \(\tau_i < 1\), but there are many lines,
  so that \(\sum_i\tau_i>>1\) over the interval \(\Delta\nu\).
In this case, the summation does not go up to
  \(N_{\max}\) either, but is truncated at
  the line with the first number \(k\), such that
  \(\sum_i^k\tau_i>1\) (i.e., \((k-N_\nu)\) terms enter into the sum,
  because the summation begins with line \(N_\nu\)).
We now obtain the following estimate:
\begin{equation} \label{eq:ctnuweek}
  \chi_{\exp}^{-1}(\nu)\approx
  ct{\frac{\nu_k - \nu}{\nu}} \approx
  ct {\frac{(k-N_\nu)\Delta\nu }{N_{\rm weak}\nu}} \; .
\end{equation}
However, \((k-N_\nu)\sim 1/ \langle \tau \rangle\),
  where \(\langle \tau \rangle\) is the mean optical depth
  of weak lines in the interval \(\Delta\nu\);
  then, \(N_{\rm weak}\langle \tau \rangle\)
  is the total optical depth of weak lines in this interval.
We obtain the same expression for weak lines from~(\ref{eq:epopac})
  by substituting \(\tau_j\) for \(1-\exp\left[-\tau_j\right]\).
Thus, both methods for weak lines yield the same result
  that is reduced simply to the summation
  of the extinction coefficients in lines
  (\(ct\) is canceled out from the definition of \(\tau_i\)),
  just as in the case of a medium at rest.

Thus, the simple heuristic \texttt{E} approximation~(\ref{eq:epopac})
  correctly conveys the limiting cases of the rigorously derived
  approximate expression~(\ref{eq:ctnupr}),
  and it may well be used in practice.
When deriving their approximation,
  the authors of \texttt{E} did not have the rigorous derivation
  done in this section,
  but they tested their recipe by a comparison with
  a rigorous numerical calculation with
  a large number of lines and obtained satisfactory agreement.
The same formula was derived even earlier by
  \citet{FriendCastor1983}
  based on the Poisson distribution of line strengths
  in a frequency interval.
The parameter \(s\) vanished in all of the last approximate
  formulas -- recall that this occurred only because
  we assumed the condition \(s_i(\delta\nu_i/\nu)<<1\).
This case is particularly important for practice,
  because the role of lines is particularly great if it is fulfilled.
Since the parameter \(s\) drops out in this case,
  the cases of both strong and weak lines considered
  by us can be described by our formulas and by Eq.~(\ref{eq:epopac})
  even at a continuum mean free path exceeding the SN ejecta sizes.
In this case, the diffusion approximation can be completely
  maintained by the forest of spectral lines.
Such a situation actually comes already near
  the maximum light of SNe~I (both Ia and Ib).

In the case of weak lines, our results also coincide with those from
  \citet{WagonerPerezVasu1991}.
However, for strong lines, when the dependences of the result
  on letter parameters coincide,
  we obtain disagreement with this paper by a factor of~2, because
  \citet{WagonerPerezVasu1991}
  adopted the definition of \(\chi_{\exp}\)
  for the transfer equation and not for the flux, as in our case.


\section*{Comparison Of The Opacities For Different
  Methods Of Averaging Over A Frequency Interval}

In the previous section we showed that the mean opacity
  derived from Eq.~(\ref{eq:epopac})
  agrees with its values inferred from~(\ref{eq:ctnupr})
  only in the limiting cases of strong and weak lines.
In the real case, a mixture of strong and weak lines,
  we have to resort to numerical calculations.

We will compare the expansion opacities derived
  for the \texttt{E} and \texttt{H} cases from
  Eq.~(\ref{eq:epopac}) and (\ref{eq:chimean}), respectively.
The averaging intervals are obtained by dividing the
  wavelength range from \(\log\,1\)~{\AA} to \(\log\,50\,000\)~{\AA}
  into 100~bins uniformly in common logarithm.
In Eq.~(\ref{eq:chimean}) \(\chi_{\exp}^{-1}\) is calculated
  from Eq.~(\ref{eq:sumfullpr}) at \(s < 30\).
For large values of this parameter we apply
  Eq.(\ref{eq:sumfullpr1}) derived in
  \citet{Blinnikov1996}
  by assuming the continuum opacity to be constant:
\makeatletter
\if@twocolumn
\begin{strip}
\fi
\makeatother
\begin{equation} \label{eq:sumfullpr1}
  \begin{multlined}
    \chi_{\exp}^{-1}(\nu) =
      \chi_{c}^{-1}\Bigl\{
        1 - \exp \Bigl[-s\Bigl(1-\frac{\nu}{\nu_{\max}}\Bigr)-
          \sum_{j=N_{\nu}}^{N_{\max}} \tau_{j} \frac{\nu}{\nu_{j}}\Bigr] - \\
          -\sum_{i=N_{\nu}}^{N_{\max }}\bigl(1-e^{-\tau_{i} \nu / \nu_{i}}\bigr)
        \exp \Bigl[-s\Bigl(1-\frac{\nu}{\nu_{i}}\Bigr)-\sum_{j=N_{\nu} + 1}^{i} \tau_{j - 1} \frac{\nu}{\nu_{j - 1}}\Bigr]
      \Bigr\} \; .
  \end{multlined}
\end{equation}
\makeatletter
\if@twocolumn
\end{strip}
\fi
\makeatother
The application of~(\ref{eq:sumfullpr1}) is related to
  the greater numerical stability of this formula at larger \(s\).
This is physically justified by the fact that larger \(s\) correspond
  to a smaller expansion opacity effect,
  because the nonthermal broadening of spectral lines
  due to the motion of the entire envelope is \(\nu/s\).
In this case, the region from which the continuum
  radiation comes to a given point is relatively small
  and a forest of lines is formed against
  the background of a constant continuum
  in numerical wavelength bins.

\begin{figure*}[!ht]
  \begin{minipage}[b]{1.0\textwidth}\centering
    \includegraphics[width=0.8\textwidth]{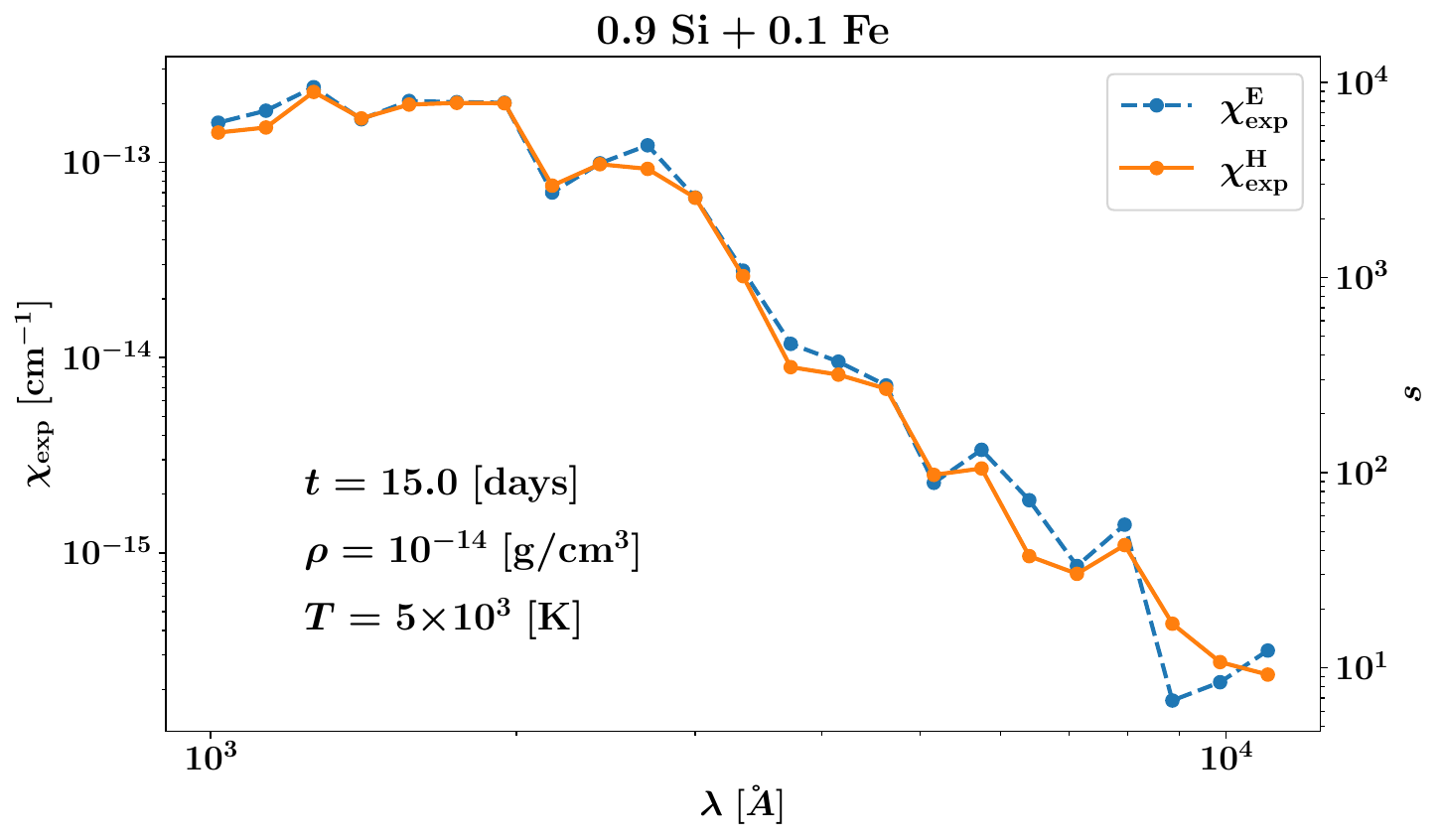}
  \end{minipage}
  \caption{
    Comparison of the ejecta opacities averaged by
      different methods over the computational cells of the wavelength grid.
    The blue curve ($\chi_{\rm exp}^{\rm E}$) corresponds
      to the \texttt{E} approximation;
      the orange curve ($\chi_{\rm exp}^{\rm H}$)
      corresponds to the Blinnikov approximation.
    The computations were performed for matter composed of
      90\% silicon and 10\% iron by mass.
    The velocity gradient corresponds to a free expansion for 15~days
      since explosion.
    The density and the temperature for which
      the computations were performed are indicated in the figure.
    All parameters roughly correspond to the SN~Ia ejecta layers
      responsible for the generation of radiation before maximum light.
  }
  \label{fig:expansion_opacity_0.9Si_0.1_Fe}
\end{figure*}
\begin{figure*}[!ht]
  \begin{minipage}[b][][b]{1.0\linewidth}\centering
    \includegraphics[scale=0.6]{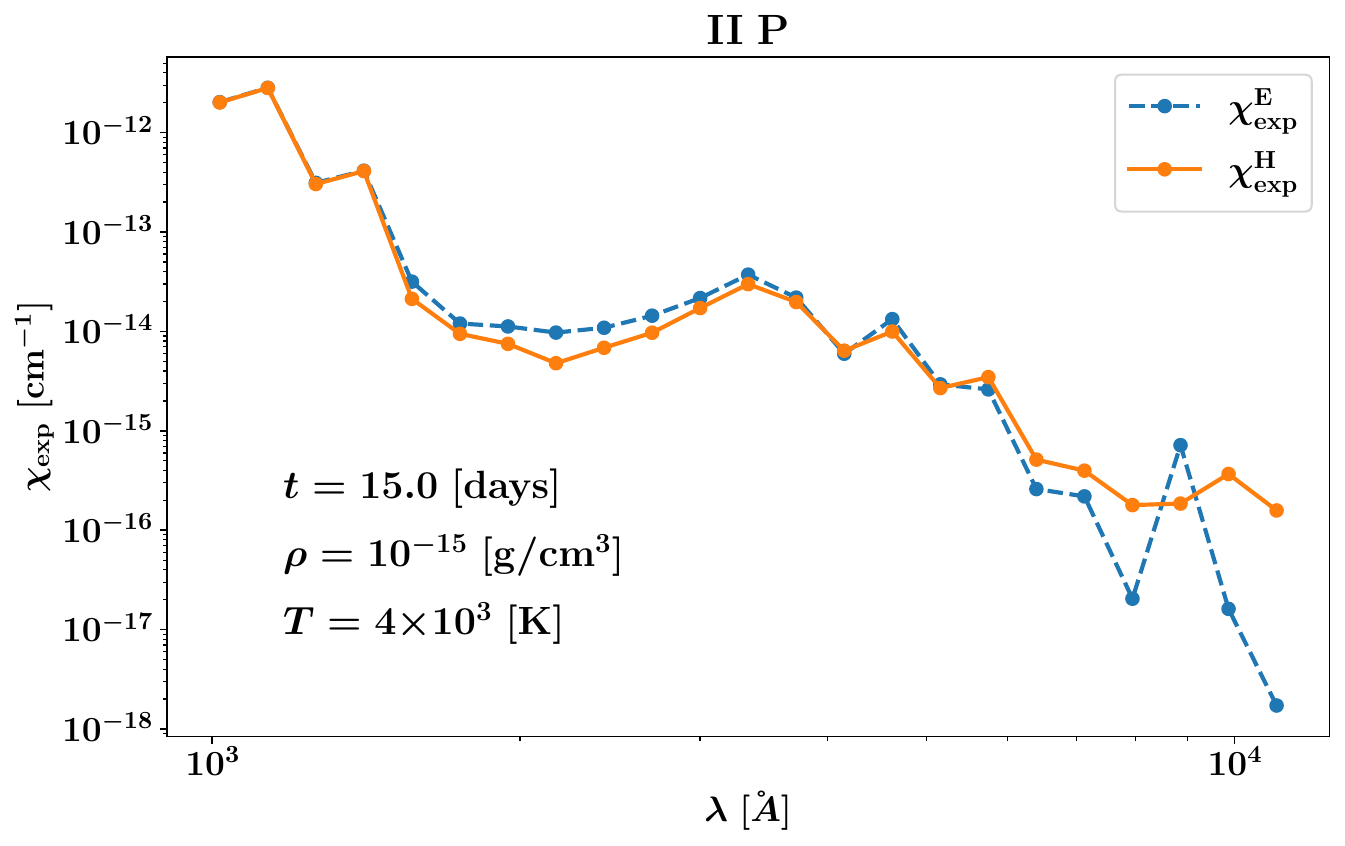} \\
  \end{minipage}
  \begin{minipage}[b][][b]{1.0\linewidth}\centering
    \includegraphics[scale=0.6]{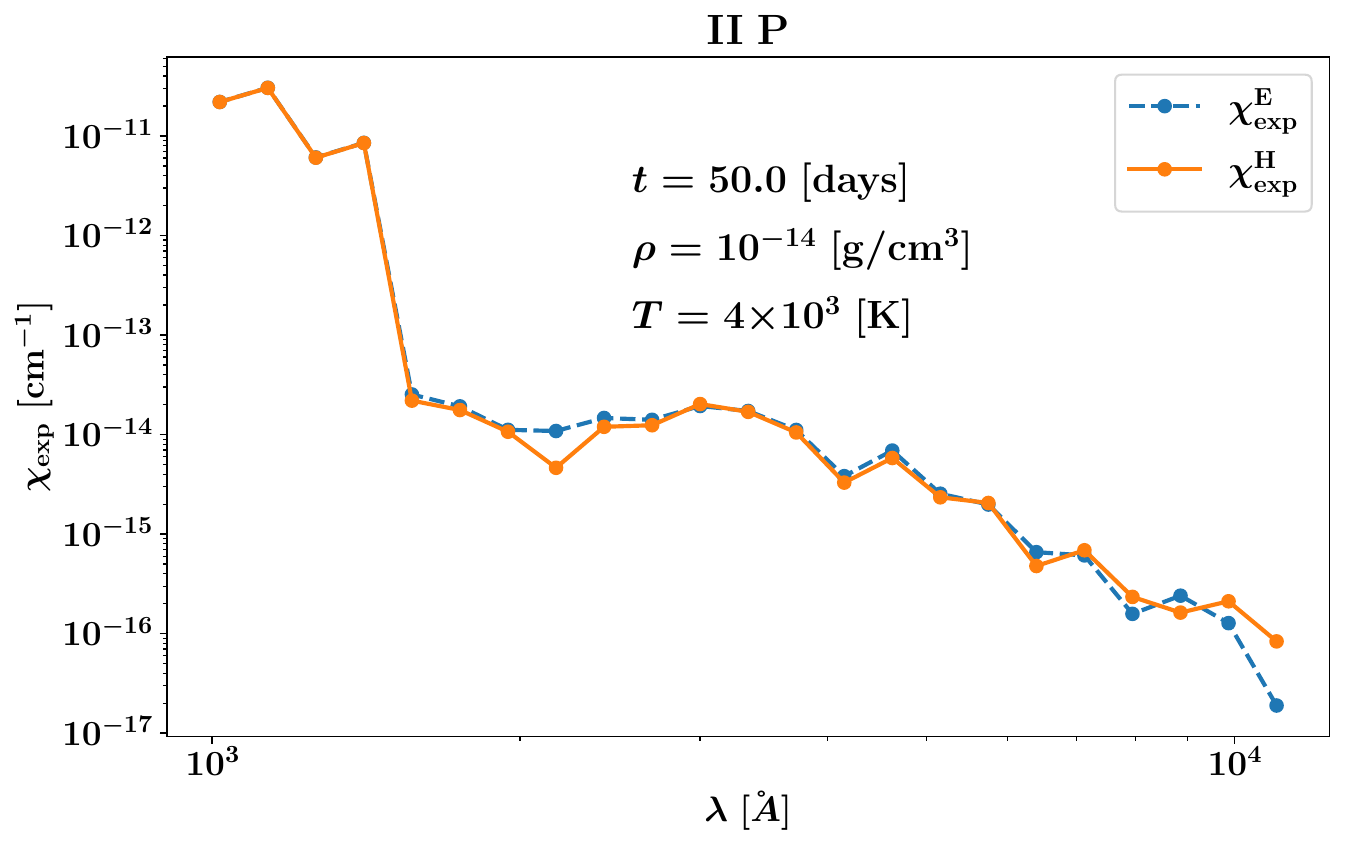} \\
  \end{minipage}
  \caption{
    Same as Fig.~\ref{fig:expansion_opacity_0.9Si_0.1_Fe},
      but for SN~IIP ejecta.
    The computations were performed for the solar chemical composition.
    The upper graph is plotted for more rarefied matter
      with a steeper velocity gradient;
      the lower one is plotted for denser matter with
      a shallower velocity gradient,
      roughly corresponding to the layers generating the SN~IIP
      radiation during the first maximum (top) and on the plateau (bottom).
    }
  \label{fig:expansion_opacity_IIP}
\end{figure*}

Figures~\ref{fig:expansion_opacity_0.9Si_0.1_Fe}
  and \ref{fig:expansion_opacity_IIP}
  present the computations of the mean opacities
  in accordance with the \texttt{E}
  \citep{FriendCastor1983, EastmanPinto1993}
  and \texttt{H}
  \citep{Blinnikov1996}
  approaches.
In Fig.~\ref{fig:expansion_opacity_0.9Si_0.1_Fe} the region
  for the W7 SN~Ia model
  \citep{NomotoThielemannYokoi1984}
  was chosen for a mixture of silicon and iron 10~days after explosion.
The \texttt{H} and \texttt{E} opacities in the visible
  and infrared ranges are seen to differ noticeably,
  which can lead to a redistribution of fluxes
  between different spectral ranges and
  to a change in the broadband light curves.

In turn, for the parameters typical of an
  SN~IIP in Fig.~\ref{fig:expansion_opacity_IIP}
  we compared the behaviors of the two approaches
  for two times (15 and 50~days after explosion)
  in regions close to the thermalization ones.
It can be seen that within the first days after
  explosion the \texttt{H} and \texttt{E} opacities
  differ more dramatically than at later stages.
Hence we can draw the preliminary conclusion that the fluxes
  will be particularly different before the light curve reaches the plateau.
This will be illustrated in the next section with specific models.


\section*{Influence Of The Opacity Averaging Method On Sn Light Curves}
  \label{sec:comparisons_light_curves}

The SN light curves were computed with the \texttt{STELLA} code
  \citep{BlinnikovEastmanBartunovEtal1998, BlinnikovRopkeSorokinaEtal2006}.
The standard procedure of this code was used
  to compute and average the opacities on a frequency grid in the
  \texttt{E} approximation.
An additional procedure was developed for
  the computations in the \texttt{H} approximation.
So far the formula from
  \citet{Blinnikov1996}
  has been used only in the \texttt{CRAB} code
  \citep{Utrobin2004}
  to compute the opacity averaged over the entire spectrum.
In our case, an averaging in narrower frequency intervals
  (\(\sim\)100~\AA) was required,
  because \texttt{STELLA} computes the radiative transfer
  on a grid of 100--1000 frequency intervals.

Using the above two opacity approximations,
  we computed the light curves for two types of supernovae,
  SNe~Ia and SNe~IIP.
The SN~Ia ejecta are composed mostly of metals and
  their opacity is determined mainly by the spectral lines,
  while the SN~IIP ejecta are mostly hydrogen ones,
  but the admixtures also play an important role.
Therefore, it should be understood how important
  the role of the ejecta expansion opacity
  averaging method for each type of SN is in modeling its radiation.

As an SN~Ia model we chose the classical W7 model
  \citep{NomotoThielemannYokoi1984}.
The initial model for an SN~IIP was constructed artificially.
This is a star of mass 15~\MS\ and radius 600~\RS\
  in equilibrium whose explosion is produced
  through the injection of thermal energy \(4\times 10^{50}\)~erg
  at the center followed by the formation of 0.15~\MS\
  radioactive nickel as a result of its explosion.
The presupernova is surrounded by an envelope
  of mass 0.03~\MS\ with a density
  \(\rho \propto r^{-2}\) up to a distance of 3600~\RS.
Such an envelope helps to explain the first maximum
  observed in many SNe~IIP in the light curve
  before it reaches the plateau.

\begin{figure*}[!ht]
  \begin{minipage}[b]{1.0\textwidth}\centering
    \includegraphics[width=0.8\textwidth]{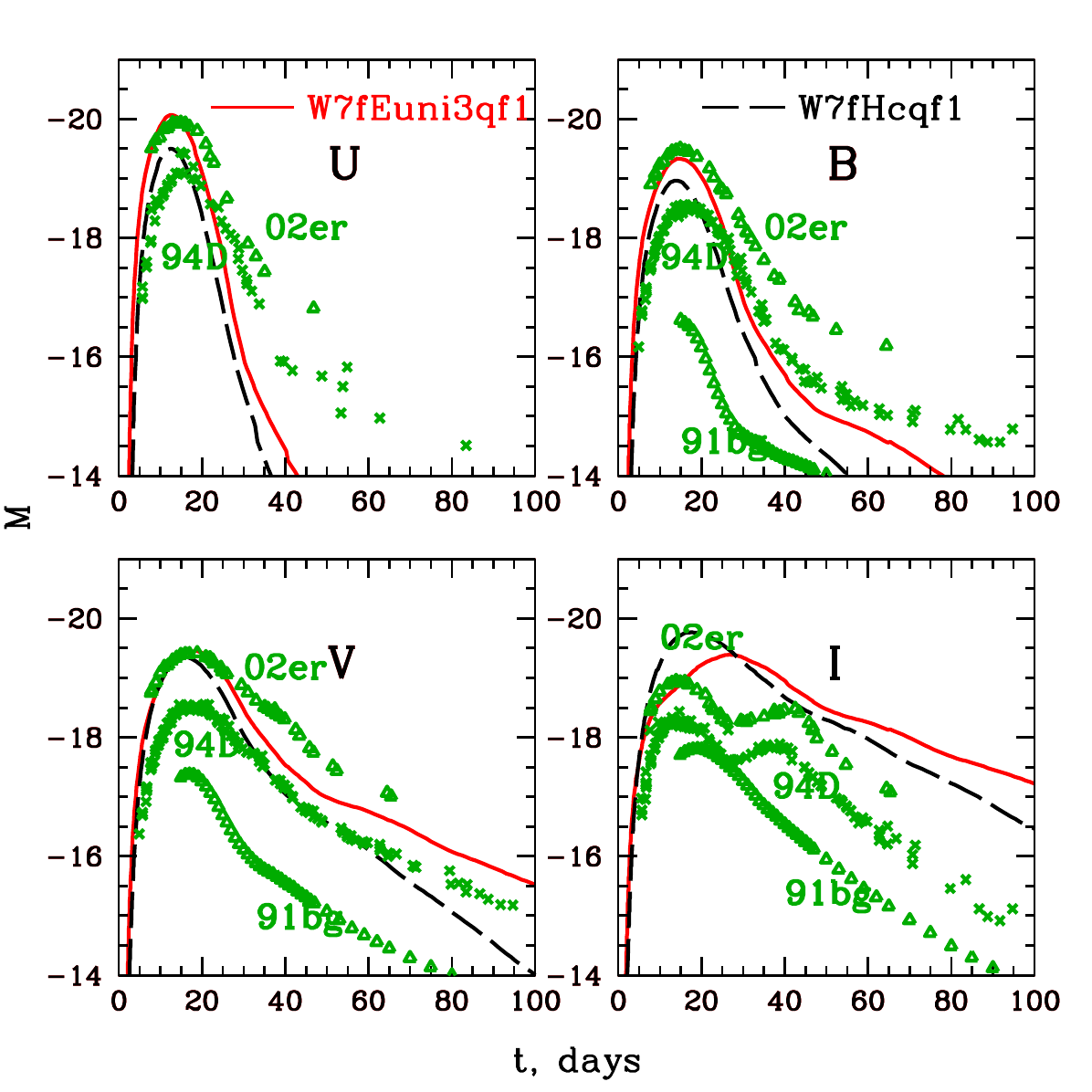}
  \end{minipage}
  \caption{
    The broadband light curves for the W7 (SN~Ia) model
      computed in the \texttt{H} (dashed lines) and \texttt{E} (solid lines)
      approximations for the opacity.
    For comparison, the crosses and triangles indicate
      the light curves of several observed SNe~Ia.
  }
  \label{fig:w7ubvi}
\end{figure*}
\begin{figure*}[!ht]
  \begin{minipage}[b][][b]{0.49\linewidth}\centering
    \includegraphics[scale=0.39]{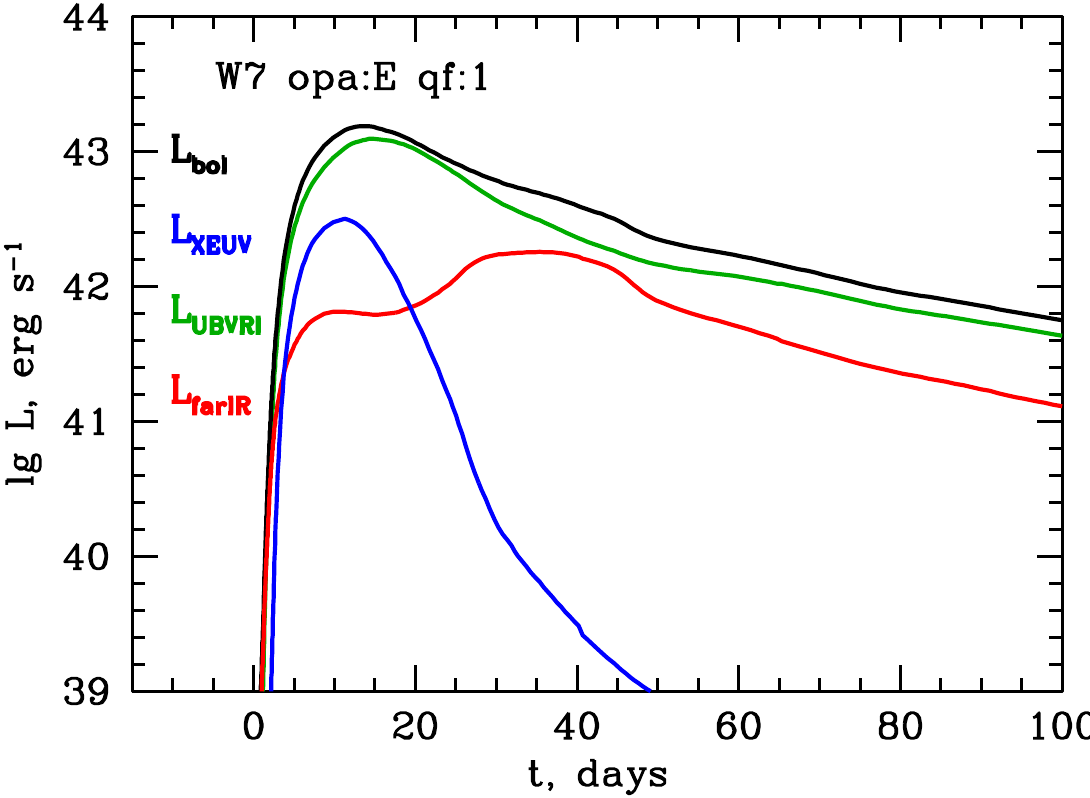} \\
  \end{minipage}
  \hfill
  \begin{minipage}[b][][b]{0.49\linewidth}\centering
    \includegraphics[scale=0.39]{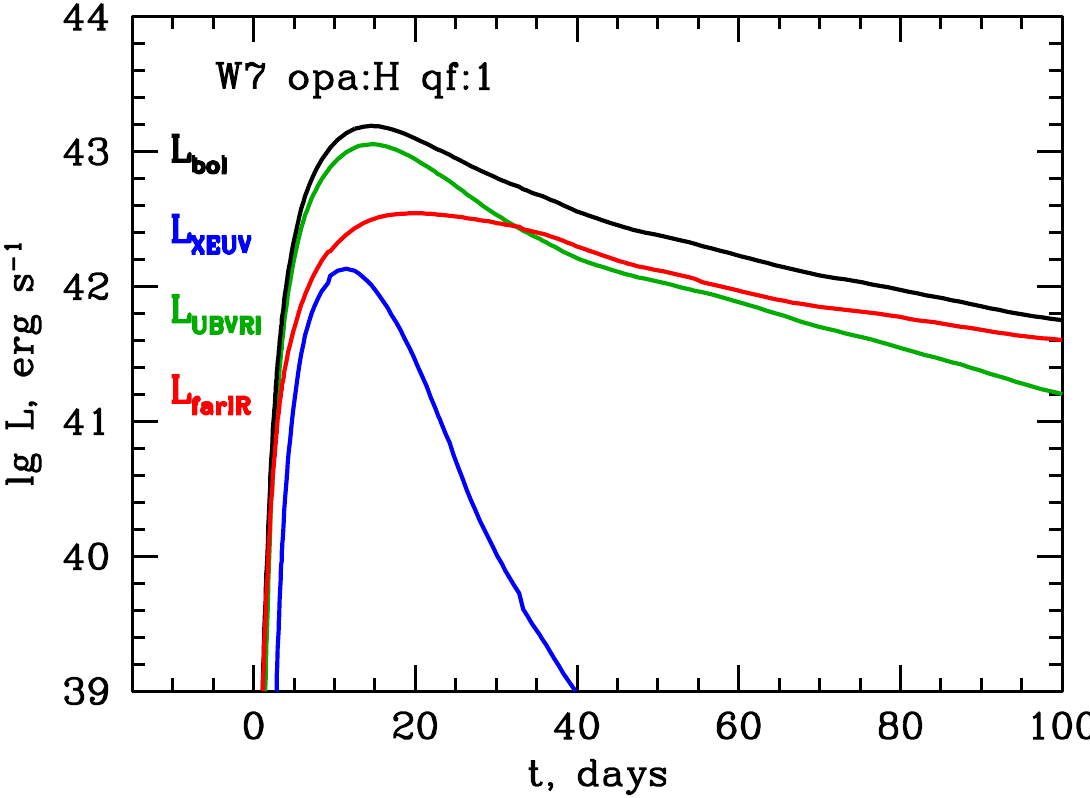} \\
  \end{minipage}
  \caption{
    The bolometric light curves (black lines) for the SN~Ia model
      computed in different approximations for the opacity:
      \texttt{E} with absorbing lines (left)
      and \texttt{H} with absorbing lines (right).
    The color lines indicate the quasi-bolometric \textit{UBVRI}
      light curve (green lines),
      the far ultraviolet blueward of the \textit{U}~band (blue lines),
      and the far-infrared radiation redward of the \textit{I}~band (red lines).
    }
  \label{fig:Iabol}
\end{figure*}
\begin{figure*}[!ht]
  \begin{minipage}[b][][b]{0.49\linewidth}\centering
    \includegraphics[scale=0.39]{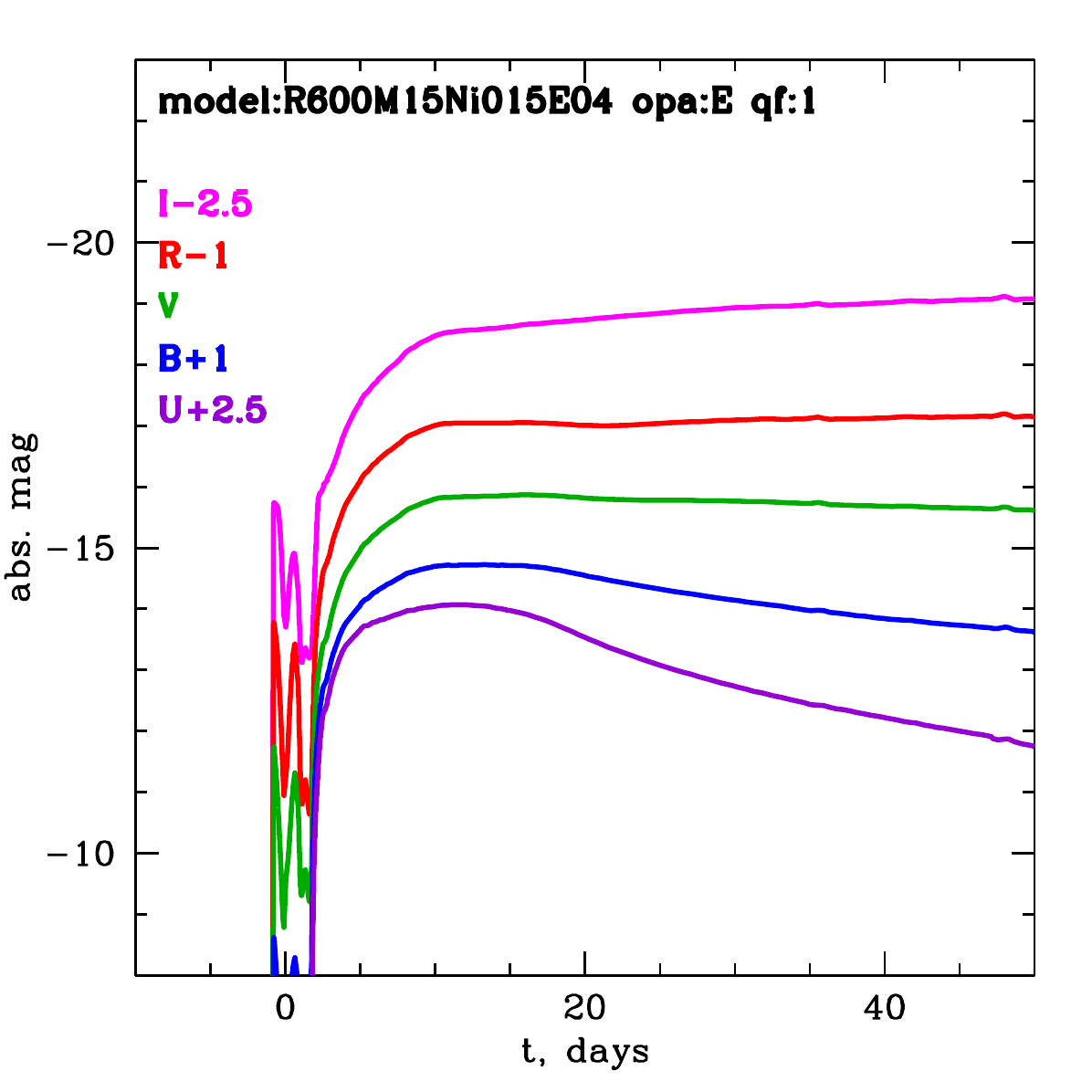} \\
  \end{minipage}
  \begin{minipage}[b][][b]{0.49\linewidth}\centering
    \includegraphics[scale=0.39]{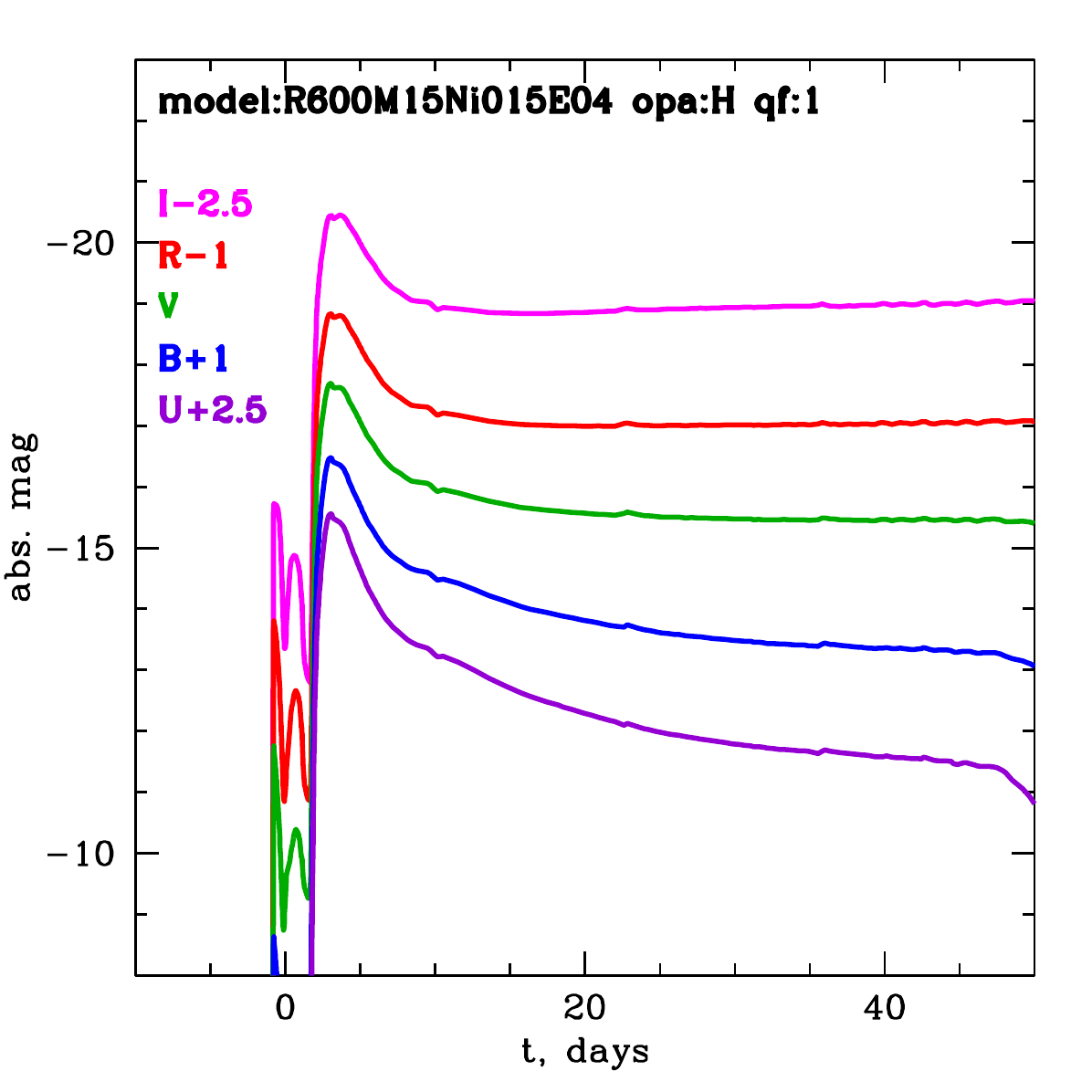} \\
  \end{minipage}
  \hfill
  \caption{
    The broadband light curves for the SN~IIP model computed
      in the \texttt{E} (left) and \texttt{H} (right) approximations
      for the opacity.
    }
  \label{fig:IIPubvri}
\end{figure*}

The \textit{UBVRI} light curves for both models
  computed in different opacity approximations
  are shown in Figs.~\ref{fig:w7ubvi} and \ref{fig:IIPubvri}.
The difference between the SN~Ia light curves
  in the visible bands is not very large.
As could be assumed from a direct comparison
  of the opacities at typical (for SN~Ia ejecta) chemical composition,
  temperature, and density
  (Fig.~\ref{fig:expansion_opacity_0.9Si_0.1_Fe}),
  in the visible range changing the approach
  to the opacity affects most strongly the \textit{I}~band.
The main differences lie in the redistribution
  of radiation between the far-ultraviolet and far-infrared
  ranges (Fig.~\ref{fig:Iabol}).
Attenuation of the hard ultraviolet flux can affect,
  for example, the degree of excitation of atomic levels
  and can change the pattern of spectral lines
  in a complete calculation including, in particular, fluorescence.

\begin{figure*}[!ht]
  \begin{minipage}[b][][b]{1.0\linewidth}\centering
    \begin{subfigure}{\linewidth}\centering
      \includegraphics[scale=0.50]{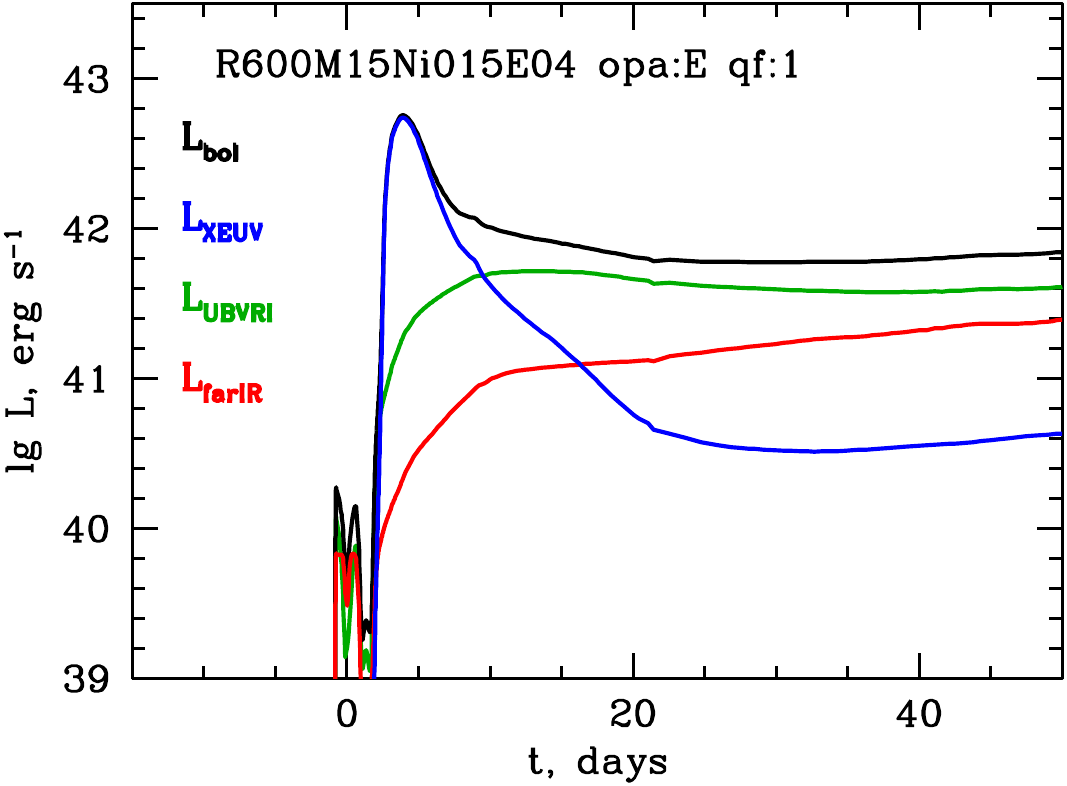}
    \end{subfigure}
    \vspace*{0.55cm}
    \begin{subfigure}{\linewidth}\centering
      \includegraphics[scale=0.50]{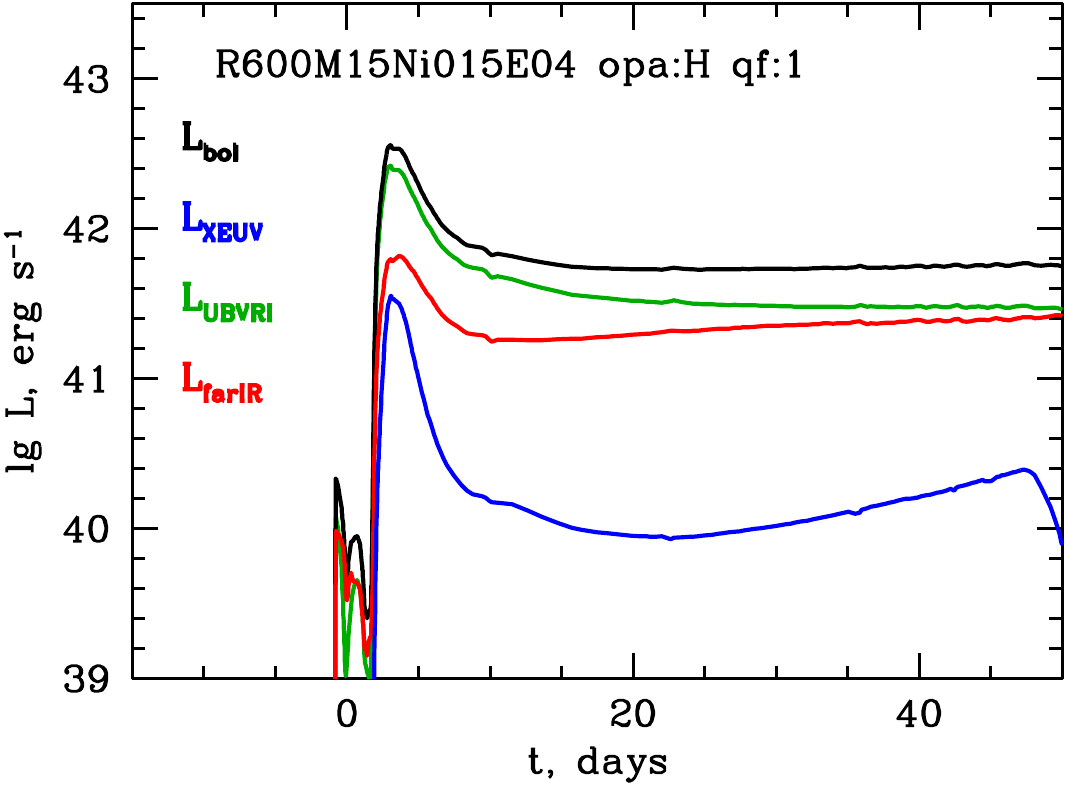}
    \end{subfigure}
    \vspace*{0.55cm}
    \begin{subfigure}{\linewidth}\centering
      \includegraphics[scale=0.50]{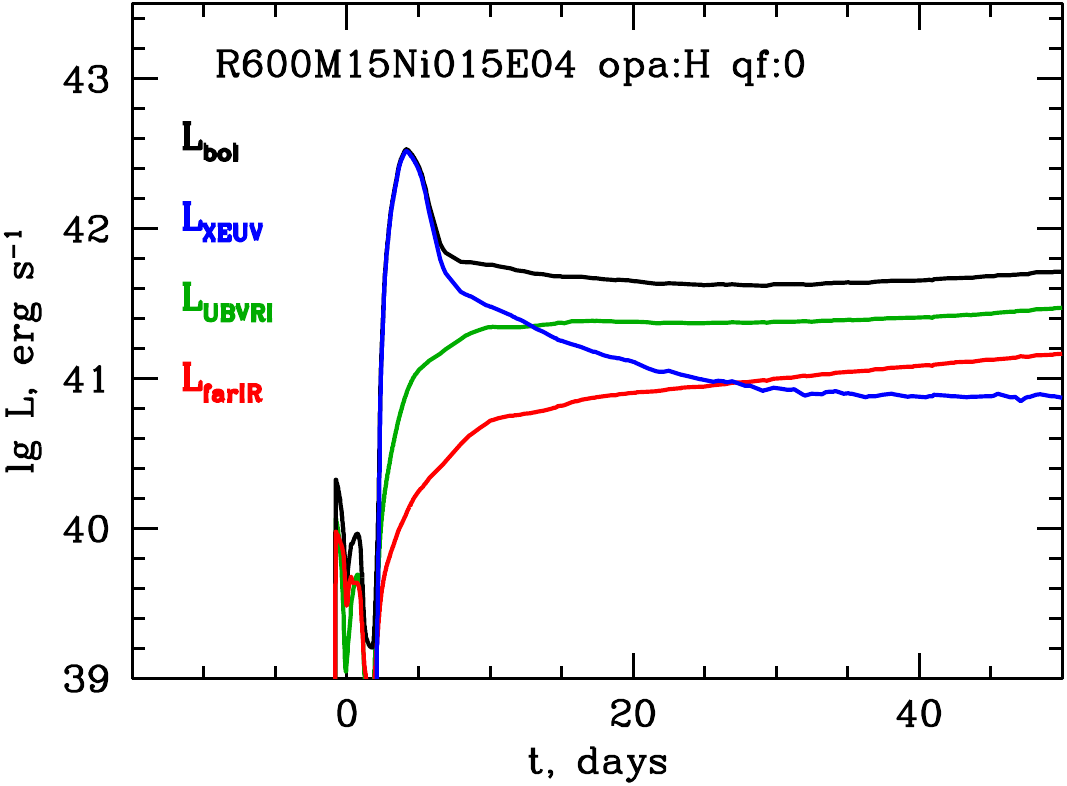}
    \end{subfigure}
  \end{minipage}
  \caption{
    The bolometric light curves for the SN~IIP model
      computed in different approximations for the opacity:
      \texttt{E} with absorbing lines (top),
      \texttt{H} with absorbing lines (center),
      and \texttt{H} with scattering lines (bottom).
    The color lines indicate the far-ultraviolet,
      quasi-bolometric,
      and far-infrared light curves, just as in Fig.~\ref{fig:Iabol}.
    }
  \label{fig:IIPbol}
\end{figure*}

In the case of SNe~IIP, the main differences in the approaches
  to the opacity manifested themselves to a greater
  extent at the initial phases of the light curve
  (the presence or absence of an initial emission
  peak before the plateau,
  see Figs.~\ref{fig:IIPubvri} and \ref{fig:IIPbol}).
Therefore, we show here only this phase, within 50~days after explosion.

The question of what contribution the spectral
  lines make to the energy exchange between radiation
  and matter and how much these lines contribute
  to the equalization of their temperatures and
  the establishment of equilibrium remains open.

\begin{figure*}[!ht]
  \begin{minipage}[b]{1.0\textwidth}\centering
    \includegraphics[width=0.75\textwidth]{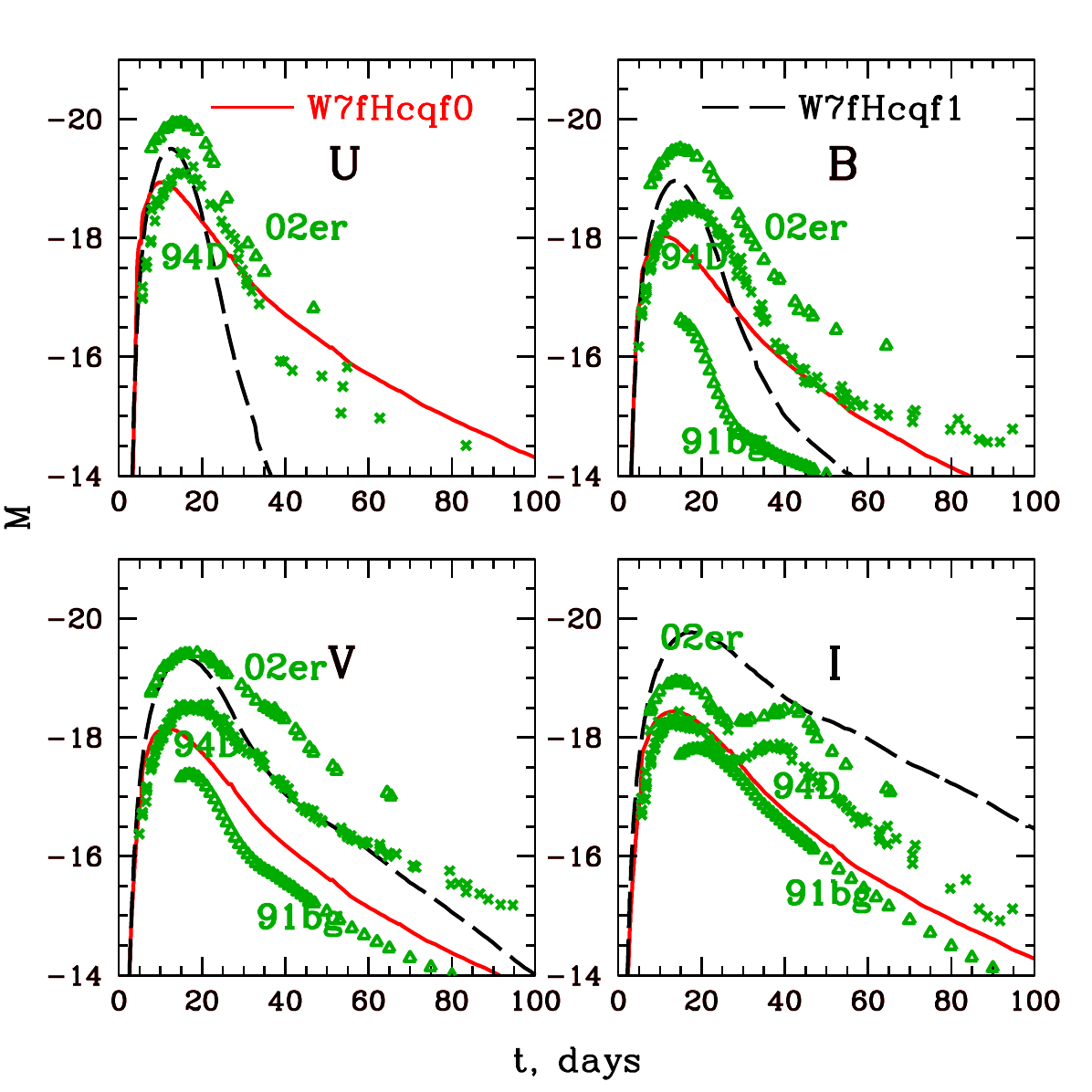}
  \end{minipage}
  \caption{
    The broadband light curves for the W7 (SN~Ia) model
      computed in the \texttt{H} approximation for the cases
      where the spectral lines are purely scattering (solid lines)
      and truly absorbing (dashed lines) ones.
    For comparison, the crosses and triangles indicate
      the light curves of several observed SNe~Ia.
  }
  \label{fig:w7ubviq01}
  \begin{minipage}[b][][b]{0.49\linewidth}\centering
    \includegraphics[scale=0.36]{magsR600Hqf1.pdf} \\
  \end{minipage}
  \hfill
  \begin{minipage}[b][][b]{0.49\linewidth}\centering
    \includegraphics[scale=0.36]{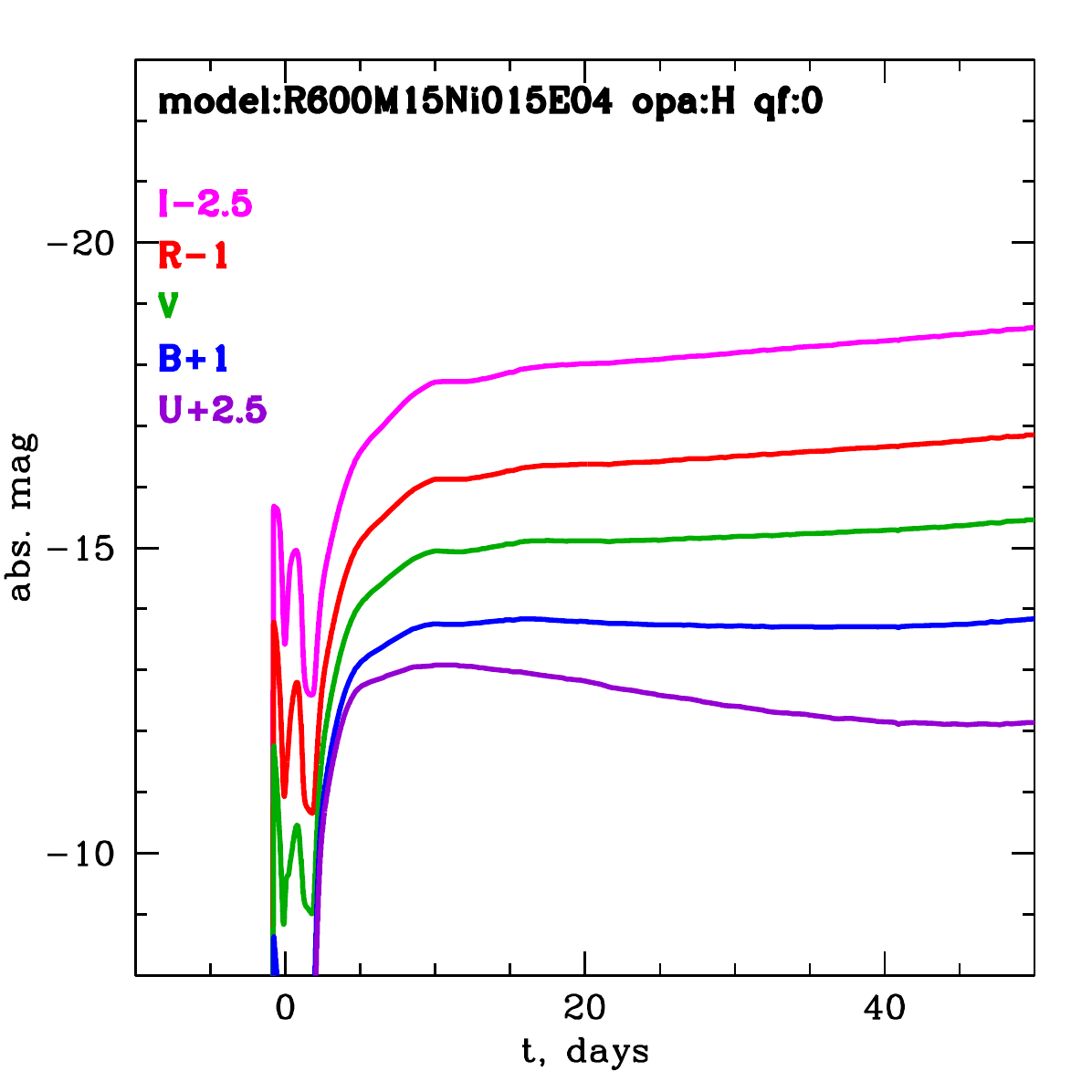} \\
  \end{minipage}
  \caption{
    The broadband light curves for the SN~IIP model
      computed in the H approximation for the cases
      where the spectral lines are truly absorbing (left)
      and purely scattering (right) ones.
    }
  \label{fig:IIPubvriqf}
\end{figure*}

As can be seen from the results of our computations
  in Figs.~\ref{fig:w7ubviq01}--\ref{fig:IIPbol},
  the height and shape of the peak changes greatly
  for different expansion opacity approximations
  and depend on the so-called \(q_f\) parameter
  that must efficiently describe the fluorescence
  \citep{EastmanPinto1993,
    BlinnikovEastmanBartunovEtal1998,
    PintoEastman2000,
    BaklanovBlinnikovPotashovEtal2013a,
    KozyrevaShinglesMironov2020}.

All of the codes to compute the light curves that
  use coherent scattering in lines neglect this important effect.
It was shown already in
  \citet{BlinnikovEastmanBartunovEtal1998}
  that the simple prescription of \(q_f = 1\),
  when all lines are purely absorbing ones,
  provides good agreement of the model light curves
  with SN observations and the \texttt{EDDINGTON} code.
Thus, for the \texttt{E} opacity a good reproduction
  for the light curves is obtained at \(q_f\) close to 1,
  when all lines are absorbing ones
  \citep[see][]{KozyrevaShinglesMironov2020},
  while for the \texttt{H} case the parameter \(q_f\) can approach zero,
  when the lines are almost purely scattering ones
  \citep{Utrobin2004, BaronHauschildtNugentEtal1996}.
The choice of an optimal parameter \(q_f\) in the \texttt{H}
  case deserves a separate study.


\section*{Discussion Of Results And Conclusions}
  \label{sec:conclusions}
\noindent

Our results show that different descriptions of
  the expansion opacity have a significant influence
  on the observed light curves of both SNe~I and II.
Comparison of \texttt{STELLA} with other codes,
  including the Monte Carlo one
  \citep{KozyrevaGilmerHirschiEtal2017, TsangGoldbergBildstenEtal2020},
  when the heuristic \texttt{E} approach is used,
  does not yet prove that the \texttt{E} approach
  is more appropriate to the problem.
After all, the \texttt{CRAB} code also shows its applicability
  to different objects and the \texttt{H} approximation
  is used there to calculate the opacity of an expanding medium.
However, it should be kept in mind that there is also
  a different aspect of the problem: fluorescence and thermalization.
Low values of the line absorption parameter should be
  taken when using the \texttt{H} approach.
In computations on powerful computers, in principle,
  no approximations like the expansion opacity could have been made,
  but so far such computations can be performed
  only in those situations where the flows are monotonic,
  there are no shock waves, and so on.


\section*{Acknowledgments}

We are grateful to V.P.~Utrobin who shared the experience
  of his work with an \texttt{H}--type approximation
  in the \texttt{CRAB} code and
  to the anonymous referee for important remarks.


\section*{Funding}

This study was supported by
  the Russian Foundation for Basic Research
  (project no.~19-02-00567).


\bibliographystyle{unsrtnat}
\bibliography{expOpacArxiv}

\end{document}